\documentclass[superscriptaddress,aps,preprintnumbers,amsmath,showpacs,amssymb,prd,nofootinbib,preprint]{revtex4-1}
\pdfoutput=1
\usepackage{booktabs} 
\usepackage{array} 
\usepackage[table,xcdraw]{xcolor} 
\usepackage{amsmath,amssymb}
\usepackage{graphicx}
\usepackage{fancyhdr}
\usepackage{bm, color}
\usepackage{slashed,amsthm,amsfonts,empheq}
\usepackage[caption=false]{subfig}
\usepackage{hyperref}
\usepackage{booktabs}
\usepackage{multirow}
\usepackage[left=2cm,right=2cm,top=2cm,bottom=2cm,a4paper]{geometry}

\newcommand{\Slash}[1]{{\ooalign{\hfil#1\hfil\crcr\raise.167ex\hbox{/}}}}

\newcommand{\beq}{\begin{equation}}  \newcommand{\eeq}{\end{equation}}
\newcommand{\bef}{\begin{figure}}  \newcommand{\eef}{\end{figure}}
\newcommand{\bec}{\begin{center}}  \newcommand{\eec}{\end{center}}

\newcommand{\laq}[1]{\label{eq:#1}}  
\newcommand{\Eq}[1]{Eq.(\ref{eq:#1})}

\newcommand{\eq}[1]{(\ref{eq:#1})}
\newcommand{\Sec}[1]{Sec.\ref{chap:#1}}

\newcommand{\vev}[1]{\left\langle {#1} \right\rangle}

\newcommand{\lac}[1]{\label{chap:#1}}

\def\({\left(}
\def\){\right)}

\def\O{\mathcal{O}}

\newcommand{\AND}{~{\rm and}~}
\newcommand{\EV}{\,{\rm eV}}
\newcommand{\KEV}{\,{\rm keV}}

\newcommand{\GEV}{\,{\rm GeV}}
\newcommand{\TEV}{\,{\rm TeV}}

\def\d{\delta}

\def\f{\phi}

\def\x{\xi}

\def\L{\Lambda}
\def\F{\Phi}

\def\tl{\tilde}
\def\*{\dagger}

\begin{document}
%\begin{titlepage}
%\begin{center}
%%\allowdisplaybreaks
%\setcounter{footnote}{0}
%\setcounter{figure}{0}
%\setcounter{table}{0}
%
%\hspace{3cm}
%
%\hspace{3cm}
%
%\hspace{3cm}
%
%\hspace{3cm}

\title{
Revisiting axion dark matter with nonlinear transitions }

\author{Max Miyazaki}
\affiliation{Department of Physics, Tokyo Metropolitan University, Minami-Osawa, Hachioji-shi, Tokyo 192-0397, Japan}

\author{Yuma Narita}
\affiliation{Department of Physics, Tokyo Metropolitan University, Minami-Osawa, Hachioji-shi, Tokyo 192-0397, Japan}
\affiliation{Department of Physics, Tohoku University, Sendai, Miyagi 980-8578, Japan}

\author{Deheng Song}
\affiliation{Department of Physics, Tokyo Metropolitan University, Minami-Osawa, Hachioji-shi, Tokyo 192-0397, Japan}

\author{Nemin Yaginuma}
\affiliation{Department of Physics, Tokyo Metropolitan University, Minami-Osawa, Hachioji-shi, Tokyo 192-0397, Japan}

\author{Wen Yin}
\affiliation{Department of Physics, Tokyo Metropolitan University, Minami-Osawa, Hachioji-shi, Tokyo 192-0397, Japan}

\begin{abstract}
Recently, two of the present authors showed that even when the axion momentum is much smaller than its mass, the axion can still behave like radiation if its energy density greatly exceeds the maximum potential energy set by the cosine-type potential. As the energy density redshifts down to the potential scale, a nonlinear transition occurs, during which the axion’s adiabatic invariant is not conserved.
In this paper, we revisit the analysis of axion dark matter by incorporating the effects of this nonlinear transition through a precise study of the axion spectrum. We demonstrate that in the parameter region with a relatively small decay constant—often favored in axion search experiments—special care is required when estimating the axion abundance and spectrum.
We also highlight a scenario in which axions are produced through the stimulated decay of a modulus, a situation that may naturally arise in the string axiverse, where the nonlinear transition occurs across a wide parameter region. Furthermore, we discuss related phenomena, including QCD axion dark matter, the formation of axion clumps such as miniclusters and axion stars, gravitational wave production, and formation of primordial black holes as dark matter.

\end{abstract}

\maketitle

\section{Introduction}

Axions and axion-like particles (ALPs) are pseudo-Nambu-Goldstone bosons whose masses originate from the explicit breaking of their continuous shift symmetries (see Refs.~\cite{Jaeckel:2010ni,Ringwald:2012hr,Arias:2012az,Graham:2015ouw,Marsh:2015xka,Irastorza:2018dyq,DiLuzio:2020wdo} for comprehensive reviews). 
Minimally, an axion $\f$ has a potential described by a single cosine term,  
\beq\laq{pote}
V(\f) = m_\f^2 f_\f^2 \bigl(1 - \cos(\f/f_\f)\bigr).
\eeq
Such a potential structure is typical for axions arising from non-perturbative effects. 
The most prominent example is the QCD axion, introduced as a natural solution to the strong CP problem through the Peccei-Quinn mechanism~\cite{Peccei:1977hh,Peccei:1977ur,Weinberg:1977ma,Wilczek:1977pj}, with the potential generated by QCD non-perturbative effects. 
In string theory, one expects a large number of axions, together with moduli, often referred to collectively as the ``string axiverse''~\cite{Witten:1984dg,Svrcek:2006yi,Conlon:2006tq,Arvanitaki:2009fg,Acharya:2010zx,Cicoli:2012sz}. Alternatively axions can be obtained in various field theoretical setups~\cite{DiLuzio:2017tjx,Lee:2018yak, Ardu:2020qmo,Yin:2020dfn,Alexander:2024nvi,Lee:2024xjb}. 
In particular, one or more of these axions can serve as the dominant component of dark matter.  

Dark matter remains one of the most compelling pieces of evidence for physics beyond the Standard Model, yet its microscopic origin is still unknown. 
Recently, various mechanisms for axion dark matter production have been proposed. 
In the standard misalignment mechanism~\cite{Preskill:1982cy,Abbott:1982af,Dine:1982ah}, the axion field begins coherent oscillations once the Hubble parameter drops below its mass, and the resulting relic energy density can account for dark matter. 
However, the predicted abundance is sensitive to the initial field value as well as to the cosmological history before and during the onset of oscillations. 
For instance, if inflation lasts sufficiently long with a low Hubble scale, the distribution of the misalignment angle can be homogenized, thereby altering the predicted relic density~\cite{Graham:2018jyp,Guth:2018hsa,Ho:2019ayl,Alonso-Alvarez:2019ixv}. 
Moreover, if the axion is initially displaced close to the hilltop of its potential, the onset of oscillations is delayed and the final abundance is enhanced~\cite{Daido:2017wwb,Co:2018mho,Takahashi:2019pqf,Nakagawa:2020eeg,Narita:2023naj}.  
In addition to coherent oscillations, axions can also be produced as particles through various channels. 
Examples include stimulated emission from other fields, inflaton or Higgs decays, and thermal scatterings~\cite{Moroi:2020has,Moroi:2020bkq,Nakayama:2021avl,Yin:2023jjj,Sakurai:2024apm}, as well as explosive production through broad resonance phenomena~\cite{Co:2017mop,Harigaya:2019qnl}. 
Axions can also be produced from topological defects such as string-wall networks~\cite{Saikawa:2024bta,Kim:2024wku,Buschmann:2024bfj,Benabou:2024msj,Kim:2024dtq,Gorghetto:2018myk,Gorghetto:2020qws,Sikivie:1982qv,Vilenkin:1982ks,Harari:1987ht,Davis:1986xc,Dine:2020pds,Hindmarsh:2021zkt,Saikawa:2024bta,Nakagawa:2022wwm,Yin:2024txg,Yin:2024pri}. 
Recent studies have also highlighted that first-order phase transitions responsible for axion mass generation can lead to efficient particle production, accompanied by novel dynamics such as bubble collisions and axion wave generation~\cite{Nakagawa:2022wwm,Lee:2024oaz}. 
Altogether, these mechanisms suggest a wide range of possibilities for axion and ALP dark matter, motivating dedicated theoretical and experimental efforts to explore their cosmological and phenomenological implications (see, e.g.,~\cite{AxionLimits}).  

More recently, two of the present authors (Y.N. and W.Y.) showed in Ref.~\cite{Narita:2025jeg} by using lattice simulation that when the total energy density of axion particles $\rho_\f$ satisfies $\rho_\f > m_\f^2 f_\f^2$, the axion field effectively behaves like radiation even when its typical momentum is below the mass. 
In other words, for the axion to behave as matter, both conditions must be satisfied: 
\beq
\laq{1}
\overline{p}_\f < m_\f, \AND \rho_{\f} < m_\f^2 f_\f^2,
\eeq
where $\overline p_\f \equiv \sqrt{\vev{p_\f^2}}$ denotes the typical momentum averaged over the axion spectrum. 
The second condition may be understood that when $\rho_\f \gg m_\f^2 f_\f^2$, the potential term in the equation of motion of the axion becomes negligible. From this result, it was demonstrated that excessively delayed matter domination would be incompatible with small-scale structure constraints. 
These dynamics also lead to stringent cosmological bounds on the axion-photon coupling. 
In addition, it was pointed out that short-lived domain walls can be produced around the {\it nonlinear transition} at $\rho_\f \sim m_\f^2 f_\f^2$.  
All these phenomena ultimately arise from the distinctive property that the axion potential, \Eq{pote}, is bounded from above.

In this paper, we use the recent finding of Ref.~\cite{Narita:2025jeg} to revisit the dark matter scenario via axion particle production, taking into account the new condition \eq{1} for the axion to behave as matter. 
We show that a large portion of the parameter space studied previously must be revised. 
In particular, we point out that when the decay constant is not very large, as favored from the viewpoint of experimental searches, the nonlinear transition can occur in various scenarios, such as axion production from the stimulated decay of moduli. 
In this case, axion clumps such as miniclusters and axion stars may be formed due to the overdensity caused by the instantaneous collapse of domain walls around the nonlinear transition.  
We also discuss the application to the QCD axion, and the heavy (non  dark matter) axion, as well as primordial blackhole formation and gravitationtal wave production. 

The paper is organized as follows. In \Sec{2}, we revisit the cosmological evolution of the axion abundance by taking into account the nonlinear transition. 
We revisit dark matter production with the nonlinear transition included, and point out a simple scenario for axion production from stimulated moduli decay. 
In \Sec{3}, we study the axion spectrum precisely by using lattice simulation. We discuss minicluster formation and gravitational wave production. 
In all the sections, so far, we consider the case where $m_\f$ and $f_\f$ remain constant in time, while in \Sec{QCD axion} we examine the QCD axion where the potential is temperature dependent.  
Finally, \Sec{6} is devoted to conclusions and discussion, and possible strong gravitational wave production.
In appendix. \ref{chap:5} we discuss nonlinear transition with the axion decay constant of $10^{15-17}\GEV$ and the axion is heavy. In appendix. \ref{chap:52}, we show that the primordial blackhole formation can occur due to the nonlinear transition of non-dark matter heavy axion.

\section{Particle axion production with nonlinear transition}\lac{2}

We now discuss the axion abundance in the context of particle production.
As demonstrated in Ref.~\cite{Narita:2025jeg}, it is crucial to properly account for the epoch at which the axion, which initially behaves as radiation, undergoes a transition to matter according to condition \Eq{1}.

One can consider the following two situations:
\vspace{-1mm}
\begin{itemize}
    \setlength{\itemsep}{0.5mm}
    \setlength{\parskip}{0.5mm}
    \item[A.] When the typical momentum of the axion, $\overline{p}_\phi$, is comparable to its mass $m_\phi$, the total axion energy density is smaller than the potential height, i.e.\ $\rho_\phi < m_\phi^2 f_\phi^2$.
    \item[B.] Even when the typical momentum is smaller than the mass, the energy density remains larger than the height of the potential, i.e. \Eq{1}.
\end{itemize}
\vspace{-1mm}
In the former case, we can estimate the axion abundance from the conservation of the number density. 
However, in the latter case, this is no longer valid. 
For simplicity, we consider axion production during the radiation-dominated era. 

\subsection{Review on particle axion abundance with comoving number conservation.}
\label{subsec:A}
Let us review the axion dark matter scenario with the particle production. The estimation in this section is valid only under the conservation of the comoving number and is therefore restricted to situations where no nonlinear transition occurs.

The axion can be produced through various mechanisms, such as the decay of other particles or topological defects, with a typical axion momentum $\overline{p}_{\phi}$ and an axion number density $n_\phi$ at a certain cosmic temperature $T_{\rm prod}$.  
{After production, the axion momentum redshifts as $\overline{p}_{\phi} \propto a^{-1}$, the number density as $n_\phi \propto a^{-3}$, and the cosmic temperature as $T \propto a^{-1}$, where $a$ is the scale factor.}
During this cooling of the Universe, we consider
\beq
\label{eq: cond1}
\boxed{\rho_{\f}< m_\f^2  f_\f^2 \text{ when }\overline p_\f \sim m_\f (>H).}
\eeq
Here, {during the radiation dominated era}, the Hubble parameter is given by $H \approx \sqrt{ \frac{g_{\star}(T)\pi^2 T^4}{90M_{\rm pl}^2}}$ {, where $g_\star(T)$ denotes the effective relativistic degrees of freedom for energy density and $M_\text{pl}$ is the reduced Planck mass}. 
Moreover, we assume that when the typical momentum is comparable to the mass scale, it is larger than $H$ so that we can neglect the Hubble friction for the mode evolution. 
In this case, the comoving axion number is conserved until the present universe.

Then, the abundance of the axion particles is usually estimated through the following procedure.
\begin{description}
\item[A1] After particle production, no further production or annihilation of axions occurs due to the weak interaction. Thus, the comoving number density $a^3 \times n_\f$ is conserved. 
The comoving momentum $a \times \overline{p}_\phi$ is also conserved. 
Since the comoving entropy density is conserved as well, we have the conserved quantities
\beq
\laq{constants}
\frac{n_\f}{s}, \qquad \frac{\overline{p}_\f}{s^{1/3}},
\eeq
where $s=\frac{2\pi^2}{45} g_{s,\star} T^3$ is the entropy density with $g_{s,\star}(T)$ denoting the effective relativistic degrees of freedom for entropy.

\item[A2] The axion abundance is then estimated as 
\beq
\laq{omeconv}
\Omega_\phi = m_\f \frac{n_\f}{s} \frac{s_0}{\rho_c},
\eeq
where $s_0$ is the present entropy density and $\rho_{\rm c}$ is
the present critical density. 
Matching with the observed dark matter abundance $\Omega_{\rm DM} h^2= 0.12$ by using the quantity at $T_{\rm prod}$ provides a relation among $n_\f$, $m_\f$, and $T_{\rm prod}$. 
The typical axion velocity $\overline{p}_\f/m_\f \ll 1$ must also be satisfied around the epoch of matter-radiation equality, so that the free-streaming length remains sufficiently small. 
\end{description}

As an illustration of this procedure, if the axion is produced from the non-relativistic particle decay such as $\F \to \f\f$,
 we can estimate  
\begin{align}
  \Omega_\f  \sim \frac{ B \rho_\F(T_\text{prod})}{s(T_\text{prod})} \frac{2 m_\f}{m_\F} \frac{s_0}{\rho_{\rm c}}, ~~~~~\overline p_\f (T) \sim  \frac{g_{\star,s}(T)^{1/3}}{g_{\star,s}(T_{\rm prod})^{1/3}} \frac{m_\F}{2}.
  \label{abundance}
\end{align}
% As an example, if the axion is produced from the decay of a non-relativistic heavy particle such as $\F \to \f\f$, 
% the resulting abundance can be estimated as
% \begin{align}
%   \Omega_\f  \sim \left.\frac{\rho_\F B}{s}\right|_{T=T_{\rm prod}} \frac{s_0}{\rho_{\rm c}}, 
%   \qquad 
%   \overline p_\f (T) \sim  
%   \frac{g_{\star,s}(T)^{1/3}}{g_{\star,s}(T_{\rm prod})^{1/3}} \frac{m_\F}{2}.
%   \label{abundance}
% \end{align}
Here, $\rho_\phi$ and $m_\phi$ denote the energy density and mass of $\phi$, respectively, and $B$ is the effective branching fraction of $\F$ decaying into an axion pair. 
Here $T_{\rm prod}$ is the decay temperature defined by the condition 
\[
H(T_{\rm prod})= \Gamma_{\F\to \f\f},
\]
with $\Gamma_{\F\to \f\f}$ being the decay rate of the process $\F\to \f\f$.

In some cases, the decay can be very efficient due to stimulated emission~\cite{Moroi:2020has, Moroi:2020bkq, Nakayama:2021avl}. 
Alternatively, axion particles can also be produced through other processes, such as the collapse of topological defects or broad parametric and tachyonic resonances.

\subsection{Particle axion abundance revisited -- the impact of nonlinear transition}
\label{subsec:B}

Particle production of axions has usually been studied in detail, whereas the subsequent evolution has not been simulated with comparable precision, for example via lattice simulations, due to computational cost and the large hierarchy of relevant timescales. 
Focusing on this later evolution, it was recently shown that the axion becomes non-relativistic only when \Eq{1} is satisfied, not merely when $\overline{p}_\f < m_\f$. 
Before this condition is met, as $\rho_\f$ redshifts, the axion behaves like radiation even if its typical momentum is non-relativistic. 
Importantly, when $\rho_\f \sim m_\f^2 f_\f^2$, it is the comoving energy density rather than the comoving number density that is conserved. 
{\it The conservation of comoving number density, assumed in step $\bf A1$ of the previous subsection, is violated} due to the nonlinear evolution of the system. 
As a consequence, the well-known estimate, i.e.\ the discussion in step $\bf A2$, no longer holds. 

% \YNC{I propose to relabel step.1-3 as step.A1-A3, and step.2', 3' as step.B1-B3. Moreover, we assume Eq.~\eqref{eq:cond2} in step B1.}
% \WYC{Agreed}

% In this case, i.e. 
{Let us consider the case where}
\beq
\laq{cond2}
\boxed{\rho_{\f}> m_\f^2  f_\f^2 \text{ when }\overline p_\f \sim m_\f (> H)}\, 
\eeq
% the steps $\bf 2$ and $\bf 3$ should be reestimated. 
% Following the radiation scaling of the component, this condition can be also written in the form
%\beq
%\laq{cond3}
%\boxed{(H<)\overline p_\f <m_\f,\text{ when } \rho_{\f}= m_\f^2  f_\f^2},
%\eeq
%by focusing on the axion matter-radiation transition. 
 To address the parameter region {consistent with the observed dark matter abundance}, {we define the transition temperature $T_\text{tr}$} by
\beq
m_\f^2 f_\f^2 = \rho_\f(T_\text{tr}).
% = s^{4/3}(T_{\rm tr}) \frac{\rho_\f(T_{\rm prod})}{s^{4/3}(T_{\rm prod})},
\eeq
% \YNC{It is a minor issue, but I think this equation is not satisfied when $g_*(T) \neq g_{*s}(T)$ for $T\lesssim 0.5 \MEV$ due to $e^+ e^-$ annihilation.}
% \YNC{I did not understand this equation, or that $\rho_\phi/s^{4/3}$ is conserved. I guess $g_{*s}(T)$ is temperature-dependent and $\rho_\phi^{1/4} \propto s^{1/3}$ is not correct.}
% \YNC{We should define $\rho_\phi$ and $s$ as functions of temp., $\rho_\phi(T)$ and $s(T)$ for simplicity.}
When axions are produced with relativistic momenta, their energy density at the production is given by $\rho_\f\simeq n_\f \overline{p}_\f$. Thereafter, the energy density redshifts as radiation.
{Then, axion is relativistic before the nonlinear transition, such that $\rho_\phi \propto a^{-4}$ and the ratio $\rho_\phi/s^{4/3}$ is conserved.}
For the cosmic temperatures slightly above $T_{\rm tr}$, the typical momentum of the axion is already smaller than $m_\f$. However, once the transition takes place, the momentum changes and the typical value at the transition is approximately
\beq\laq{ptr}
\overline{p}_\f(T_{\rm tr}) \sim m_\f.
\eeq
This behavior is illustrated in Fig.\,\ref{fig:3} by using lattice simulation.

{When Eq.~\eqref{eq:cond2} is satisfied, procedures {\bf A1} and {\bf A2} require reevaluation. Accordingly, we will estimate the axion abundance using an different method.}

\begin{description}
\item[B1] 
{For {$T_{\rm prod}>T > T_\text{tr}$}, the conserved quantities are
\begin{equation}
    \frac{\rho_\f(T)}{s^{4/3}(T)}{=\frac{\rho_\f(T_{\rm prod})}{s^{4/3}(T_{\rm prod})}} \sim \frac{m_\f^2 f_\f^2}{s^{4/3}(T_\text{tr})}, ~~~~\frac{\overline{p}_\f(T)}{s^{1/3}(T)}\sim {\frac{\overline{p}_\f(T_\text{prod})}{s^{1/3}(T_\text{prod})}}.
\end{equation}
From the first condition, the transition temperature $T_{\rm tr}$ is determined given the function of $\rho_\phi(T)$.}
{After the nonlinear transition}, axions behave as matter. Consequently, the conserved quantity relevant to the energy density is replaced by
\beq
\frac{\rho_\f(T)}{s(T)}\sim \frac{m_\f^2 f_\f^2}{s(T_\text{tr})},
\eeq
for any $T<T_{\rm tr}.$ {Although nonlinear transition alters the typical momentum,} the comoving axion momentum is conserved as $\overline{p}_\phi(T) / s^{1/3}(T) \sim m_\phi / s^{1/3}(T_\text{tr})$.
\item[B2] Thus, we obtain
\beq
\Omega_\phi^{'} \sim \frac{m_\f^2 f_\f^2}{s(T_{\rm tr})}\frac{s_0}{\rho_c}=
s^{1/3}(T_{\rm tr})\frac{\rho_\f(T_{\rm prod})}{s^{4/3}(T_{\rm prod})} \frac{s_0}{\rho_c}. 
\eeq
This is usually smaller than \Eq{omeconv}, which can be found from another formula 
$\Omega_\phi \sim  s^{1/3}(T_{m_\f}) \frac{\rho_\f(T_{\rm prod})}{s^{4/3}(T_{\rm prod})} \frac{s_0}{\rho_c}$. Here, $T_{m_\f}$ is the temperature when $\overline p_\f=m_\f$.
In other words, we get 
\begin{equation}
    \laq{3'}
    \Omega_\phi^{'} \sim \frac{s^{1/3}(T_{\rm tr})}{s^{1/3}(T_{m_\f})}\Omega_\phi
\end{equation}
 Matching with the dark matter abundance $\Omega_{\rm DM} h^2= 0.12$ gives the condition that the axion explains {the dominant dark matter}.
 In addition we need to require $\overline{p}_\f/m_\f\ll 1$ at around the matter radiation equality, which essentially gives the generic limit derived in, e.g.,~\cite{Marsh:2019bjr,Narita:2025jeg}, which directly constrains that $T_{\rm tr}< T_{0} (1+5.5\times 10^6)\approx 1.3\KEV$.
\end{description}

\begin{figure}[!t] 
    \begin{center}
        \includegraphics[width=120mm]{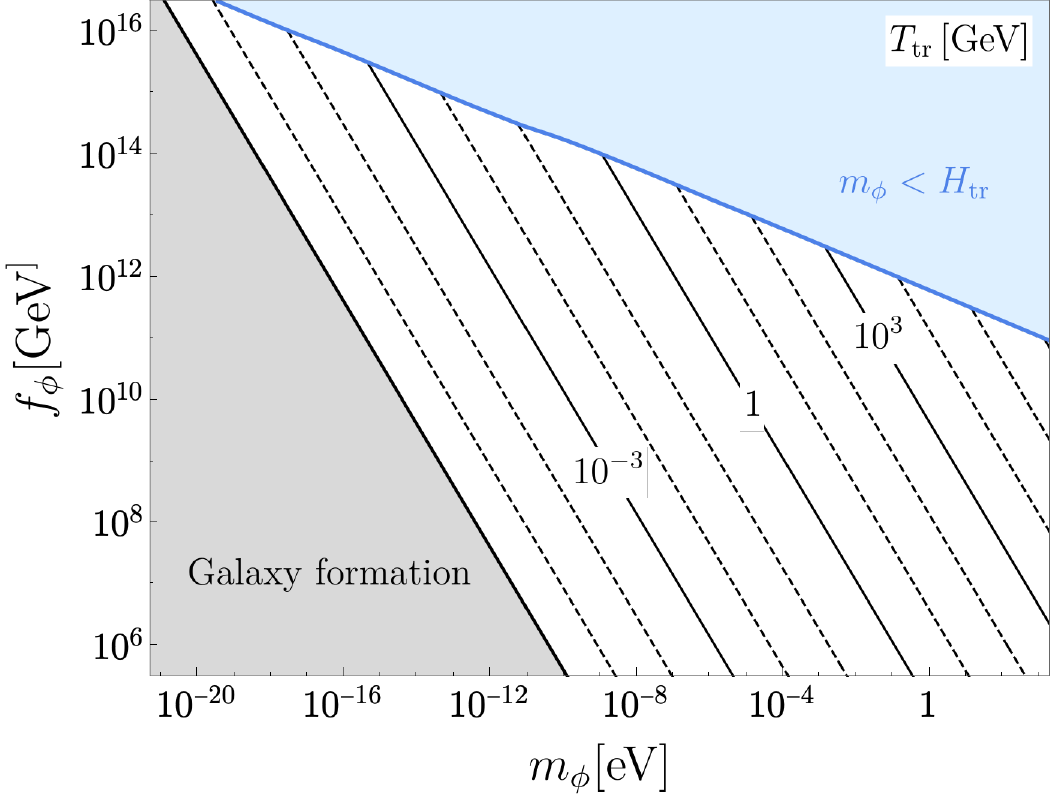}
\vspace{-5mm}
    \end{center} 
    \caption{The contour plot of the transition temperature $T_{\rm tr}$ in the $m_\f \,\text{-}\, f_\f$ plane 
    for the nonlinear transition explaining axion dark matter. The relativistic degrees of freedom are taken from \cite{Husdal:2016haj, Saikawa:2018rcs}. {The light-gray region is excluded by the generic bound from galaxy formation derived in \cite{Narita:2025jeg} (see also \cite{Marsh:2019bjr,Dror:2020zru}). The upper limit, shown in the light-blue region, corresponds to transition momentum/mass scales smaller than the Hubble parameter. 
    }}
    \label{fig:1DMfphi} 
\end{figure}

\begin{figure}[!t] 
    \begin{center}
        \includegraphics[width=120mm]{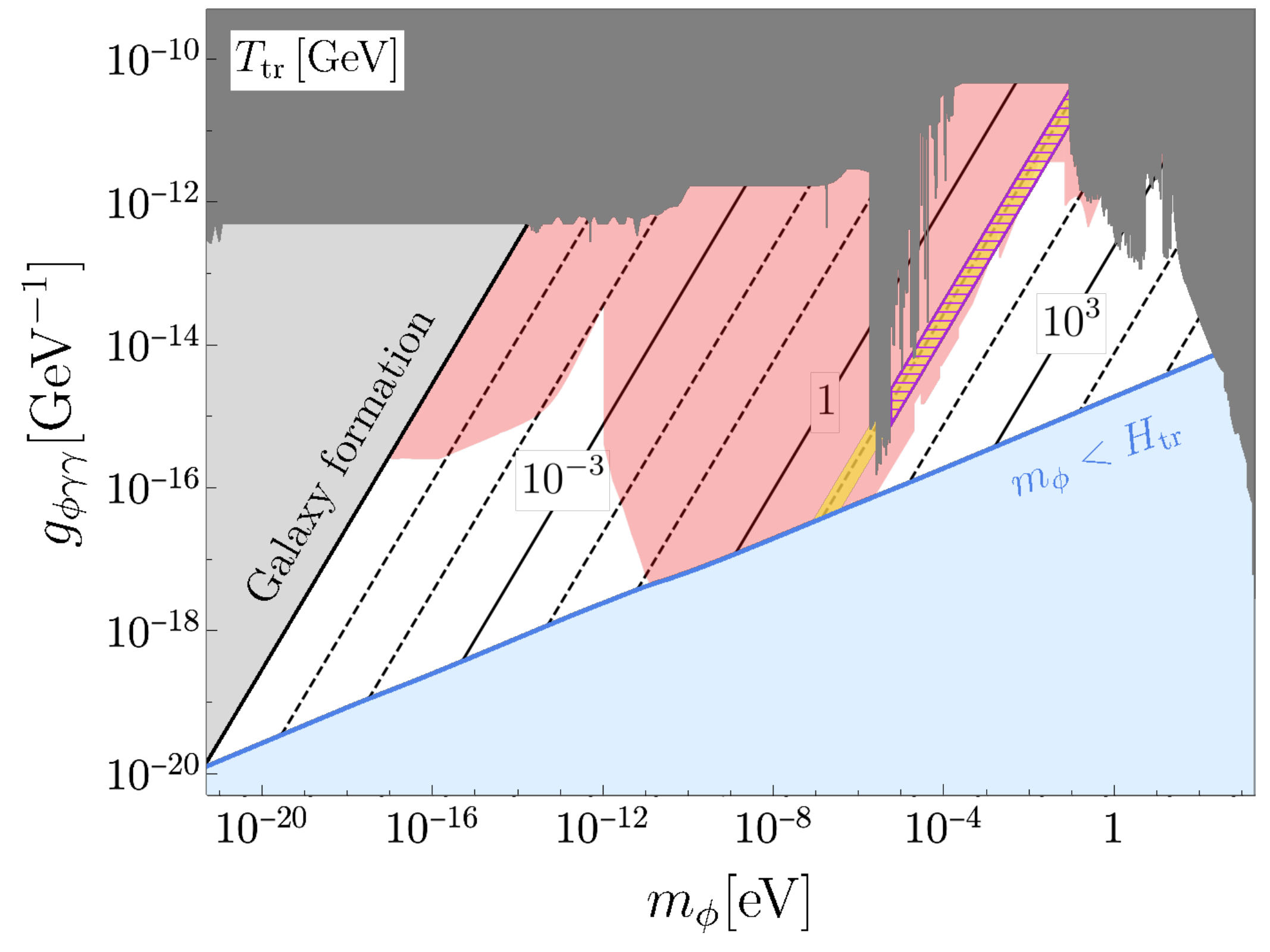}
\vspace{-5mm}
    \end{center} 
    \caption{The contour plot of the transition temperature $T_{\rm tr}$ 
    % in the $m_\f$-$f_\f$ plane 
    for the nonlinear transition explaining axion dark matter in the parameter space of the axion mass $m_\phi$ and the axion-photon coupling $g_{\phi \gamma \gamma}$. 
    % The relativistic degrees of freedom are taken from \cite{Husdal:2016haj, Saikawa:2018rcs}. 
    {Here, we take $C_\gamma=1$. The light-gray and light-blue regions correspond to those shown in Fig.~\ref{fig:1DMfphi}. The yellow band represents the QCD axion-photon coupling predicted between the KSVZ model~\cite{Kim:1979if, Shifman:1979if} and the DFSZ model~\cite{Dine:1981rt, Zhitnitsky:1980tq}. {The hatched purple region corresponds to the QCD axion prediction with a nonlinear transition, $T_{\rm tr}\sim 1.3\GEV, f_\phi \lesssim 10^{12}\GEV$, which is independent of the axion mass (see \Sec{QCD axion}).}
     The dark-gray regions indicate constraints from current observations and experiments, while translucent red regions indicate the projected sensitivity of upcoming experiments. These regions are taken from the data compiled in Ref.~\cite{AxionLimits}.}}
    \label{fig:1DM} 
\end{figure}

If $T_{\rm tr}$ is conservatively treated as a free parameter, the discussion can proceed independently of the production mechanism.
Analytically, we expect the axion abundance as 
\begin{equation}
    \label{eq:axionabundance}
    \Omega_\phi' h^2 \sim 0.12 \left( \frac{m_\phi}{10^{-7} \EV}\right)^2 \left( \frac{f_\phi}{10^{12} \GEV}\right)^2 \left( \frac{T_\text{tr}}{1 \GEV}\right)^{-3},
\end{equation}
where we have used $g_{*s} \approx 60$, and $\rho_c/s_0 = 3.644 \times 10^{-9} \GEV \, h^2$. {Fig.~\ref{fig:1DMfphi} shows the contours of the transition temperature in $m_\phi \text{-} f_\phi$ plane. The bottom light-blue region corresponds to $m_\f < H_{\rm tr}\equiv H|_{T=T_{\rm tr}}$, where our discussion so far does not apply.\footnote{This implies an even later-forming dark matter scenario, since at $T=T_{\rm tr}$ one has $\overline{p}_\f < m_\f < H_{\rm tr}$. Most of the dark matter modes are frozen by the Hubble friction.}
The light-gray region, denoted as galaxy formation, is excluded due to the excessively long free-streaming length.}

{When discussing the verifiability of axion parameters, many axion search experiments rely on its coupling to photons. We assume that the axion couples to a pair of photons with the coupling
\begin{equation}
    \mathcal{L} = -C_\gamma \frac{\alpha}{2 \pi} \frac{\phi}{f_\phi} F \tilde{F} \equiv - \frac{1}{4} g_{\phi \gamma \gamma} F \tilde{F},
\end{equation}
where $g_{\phi \gamma \gamma}$ is the axion-photon coupling constant. {In this case, the axion is called ALP.} For simplicity, we take $C_\gamma = 1$.
}Figure~\ref{fig:1DM} shows the parameter region with contours of the transition temperature consistent with the present dark matter abundance
% , shown
in the plane of $m_\phi$ versus $g_{\phi \gamma \gamma}$. 
{Outside the shaded regions, the dynamics described in this section can occur in the early Universe and remain consistent with the current constraints, if} $\overline p_\f < m_\f$ at $T=T_{\rm tr}$.\footnote{{A simple model realizing this possibility is to produce dark matter with low momentum just before $T=T_{\rm tr}$. For the heavy particle decay we can consider $m_\Phi \sim 2 m_\f$. 
Although light the stimulated emission can be enhanced with $\overline p_\f< m_\f$ at the production when $m_\F \approx 2 m_\f$ (see Eqs.~(2.6), (2.12), and (2.13) of \cite{Moroi:2020has}).}}
The viable region for axions in Fig.~\ref{fig:1DM} includes the QCD axion band; however, the QCD axion with its temperature-dependent potential gives different prediction, $T_{\rm tr}\sim 1.3\GEV, f_\phi \lesssim 10^{12}\,\GEV$
% {, in the hatched purple region,} 
which will be discussed in Sec.~\ref{chap:QCD axion}.

{Interestingly, the parameter region relevant to the dynamics discussed in this paper is typically associated with a large axion–photon coupling. In such cases, the effects can become significant, particularly in the context of future axion dark matter searches (see the red shaded region in Fig.~\ref{fig:1DM}). Therefore, one should exercise caution when studying the particle production of axion dark matter that is searched for.}

% \YNC{Should we discuss the constraint of $\Delta N_\text{eff}$ in this case for $T_\text{tr} < T_\text{BBN}$?}

\subsection{Generic estimation of axion dark matter abundance}
We emphasize that the relevance of the nonlinear transition depends on the typical momentum of the produced axion, which is assumed to be sufficiently low in Sec.~\ref{subsec:B}.
If the momentum is relatively high, the evolution discussed in Sec.~\ref{subsec:A} must also be included.  
{To evaluate the axion abundance from the particle production, one needs to determine whether the energy density is higher or lower compared to the potential height when $\overline{p}_\phi = m_\phi$. However, since axions behave as matter {only when 
% \WY{both conditions in} 
Eq.~\eq{1} is} satisfied, we should estimate both contributions and simply adapt the smaller one.}
By noting the relation \eq{3'}, we obtain the generic condition for the axion to explain the observed dark matter abundance,
\beq\laq{ODM}
\min[\Omega_\f,\Omega'_\f] = \Omega_{\rm DM}.
\eeq
% i.e., we estimate both contributions and take the smaller one to account for the dark matter abundance.
We also find that the condition for the nonlinear transition is $s(T_\text{tr})<s(T_{m_\phi})$. Using the relevant conserved quantities in steps \textbf{A1} and \textbf{B1}, one obtains the requirement that the typical momentum at particle production satisfies
\begin{equation}
    \label{eq: pcondition}
    \overline{p}_\phi(T_\text{prod}) \lesssim \sqrt{\frac{m_\phi}{f_\phi}} \rho_\phi^{1/4}(T_\text{prod}),
\end{equation}
which ensures that the nonlinear transition occurs. Consequently, the new contribution $\Omega'_\f$ becomes significant when axions are abundantly produced with sufficiently low momentum, as in Eq.~\eqref{eq: pcondition}.
As we will show, this situation naturally arises when axions are produced through modulus-stimulated decay. 

\subsection{Axion dark matter from modulus-stimulated decay}
\label{subsec:D}

As an illustrative example, let us build a model in which a {CP-even scalar $\F$ decays} as $\F \to \f\f$, producing a pair of $\f$. {$\F$ is a non-relativistic condensate.
Since we will set the natural energy scale of the couplings on the order} of the Planck scale, we may regard $\F$ as a modulus, analogous to the moduli in string theory.  
The relevant Lagrangian for axion production is 
\beq\laq{Lag}
{\cal L} \supset -(\x_2 \F^2 + \x_1 M_{\rm pl} \F) R - V(\F)+ g_{\F\f\f}\, \F \partial^\mu \f \partial_\mu \f,  V(\F)\approx \frac{m_\F^2}{2} \F^2(1+ \O(\F/M_{\rm pl}))
% +\O(\F^3) 
\eeq
Here, $m_\F \gg m_\f$, and $g_{\F\f\f}$ denotes the coupling of $\F$, which can be very small. 
We neglect the term such as $\Phi(1 - \cos(\phi/f_\phi))$, which can be generated with the non-perturbative effect relevant to the axion mass, and is suppressed by the small coupling e.g. $\L^4/M_{\rm pl}$. The parameters $\x_i \sim \O(1)$ represent non-minimal couplings.

{During inflation, $\F$ is driven to a field value $\F \simeq -\x_1 M_{\rm pl}/(2\x_2) =\O(M_{\rm pl})$ due to the Hubble induced mass squared $12\xi_2 H_{\rm inf}^2$, which can be larger than the }inflationary Hubble parameter squared $H_{\rm inf}^2$ {if $\x_2\gtrsim \O(0.1).$} Then, the modulus does not acquire the isocurvature fluctuations and we do not have isocurvature problem.  After inflation and reheating, the Hubble induced mass becomes highly suppressed and the modulus is frozen until the Hubble friction is not significant. 
When $m_\F \gtrsim H$, the field $\F$ begins to oscillate around the potential minimum. 
The corresponding number density at this epoch is estimated as
\beq
n_\F \sim \frac{m_\F}{2} M_{\rm pl}^2.
\eeq

\begin{figure}[!t] 
    \begin{center}
        \includegraphics[width=120mm]{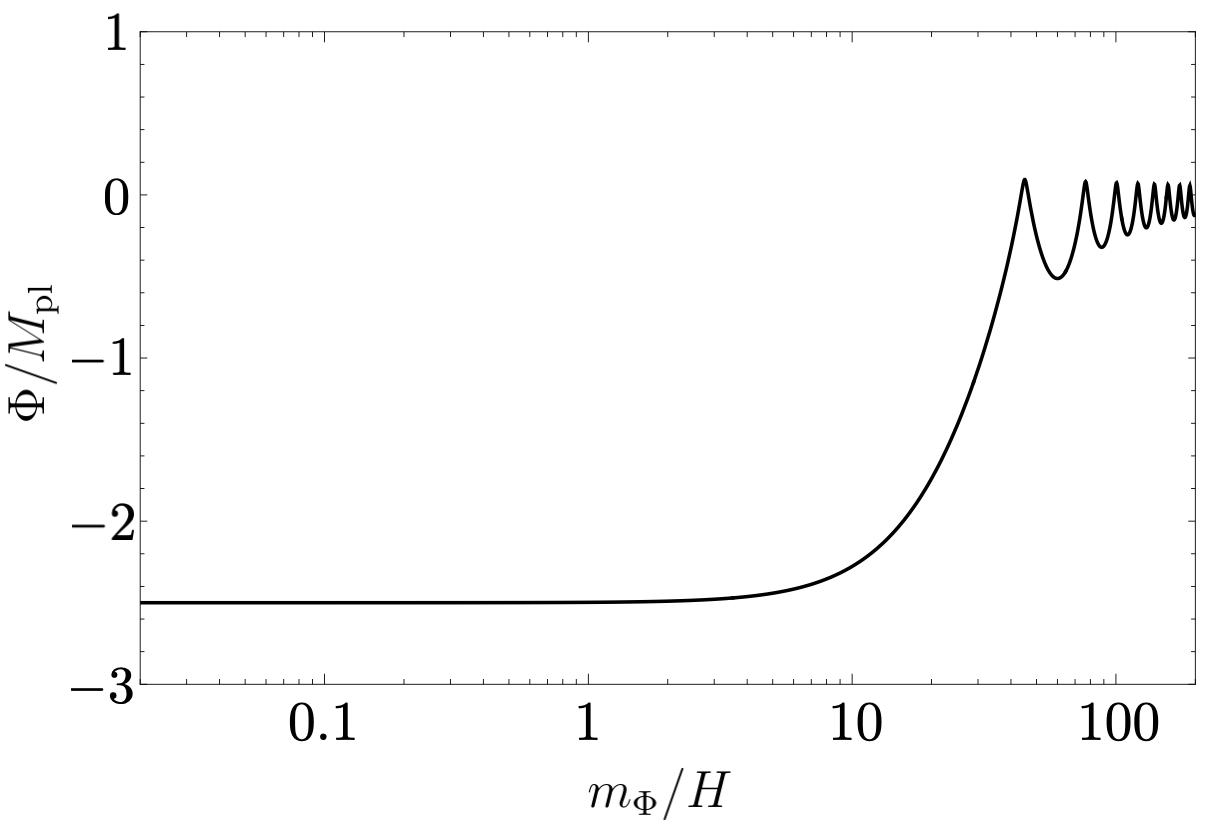}
\vspace{-5mm}
    \end{center} 
    \caption{The delay of the onset of oscillation due to the non-minimal coupling. We take $\xi_1=5,\x_2=1$.}
    \label{fig:osc} 
\end{figure}

{Note that the onset of oscillation can be delayed compared to the na\"{i}ve expectation at $m_\Phi \sim H$ due to the non-minimal coupling, which will be a condition for the stimulated emission. These couplings also modify the potential shape when $\Phi = \mathcal{O}(M_{\rm pl})$.  
In fact, after Weyl rescaling the potential effectively becomes
\beq
V_\text{eff}(\Phi) = \frac{V(\Phi)}{\left(1 - \xi_1 \frac{\Phi}{M_{\rm pl}} - \xi_2 \frac{\Phi^2}{M_{\rm pl}^2}\right)^2}.
\eeq}
{Depending on the higher-dimensional terms in $V(\Phi)$ and the values of $\xi_1, \xi_2$, one can easily obtain a potential that delays the onset of oscillations. For instance, neglecting higher-order terms and taking $\xi_1 = 5, \xi_2 = 1$, the numerical solution to the equation of motion  for $\Phi$ with the initial conditions $\dot{\F}=0, \F=-\x_1 M_{\rm pl}/(2\x_2)$ shows that oscillations begin when $m_\Phi \approx 40H$ (see Fig.\,\ref{fig:osc}). The larger the value of $|\xi_1|$, the later the onset of oscillations. This is because for large $\xi_{1,2}$, when $\Phi$ rolls down the potential, $\x_2$ gradually becomes less important than $\x_1$ term. Then for a while the $\xi_1$ term dominates the numerator of the potential, and the potential becomes approximately flat since both the numerator and denominator scale as $\Phi^2$.} 

The coupling, $g_{\F\f\f}$, is naturally obtained from the mixing between $\F$ and the PQ Higgs in the UV completion of the axion.  
To see this, consider the interaction term $A \F |\F_{\rm PQ}|^2$.  
The dimensionless mixing parameter can be estimated as 
$
\theta_{\F-\F_{\rm PQ}} \sim \frac{A}{f_\f},
$
assuming that the PQ Higgs mass scale and vacuum expectation value are both of order $f_\f$.  
For simplicity, let us focus on the region $A M_{\rm pl} < f_\f^2$, so that the field excursion of $\F$ does not affect the PQ Higgs mass and VEV.  
This condition gives the bound $\theta_{\F-\F_{\rm PQ}} < f_\f/M_{\rm pl}$.  
In this UV completion, since the PQ Higgs derivatively couples to the axion with a $1/f_\f$ scaling, the effective coupling is 
$
g_{\F\f\f} \sim \frac{\theta_{\F-\F_{\rm PQ}}}{f_\f} \;\lesssim\; \frac{1}{M_{\rm pl}}.
$ 
The decay rate is then estimated as
\beq
\Gamma_{\F \to \f\f} = \frac{g_{\F \f\f}^2}{32\pi}\, m_\F^3.
\eeq

Soon after the onset of oscillation, the number density of axions is significantly enhanced due to stimulated emission.  
By solving the Boltzmann equation analytically and neglecting the non-minimal coupling correction and the decrease in the comoving number of $\F$
(see the more precise discussion in \cite{Moroi:2020bkq}), the number density of axions soon after the onset of oscillation is given by
\begin{align}
  n_{\f} \sim \frac{m_\F^3}{32\pi^2} \left(e^{2\overline f}-1\right), 
  \qquad 
  \overline{p}_\f \simeq \frac{m_\F}{2},
\label{number}
\end{align}
where  
\beq  
\overline{f} \equiv \frac{32\pi^2 \Gamma_{\F \to \f\f} n_\F}{H m_\F^3},
\eeq
and we have taken the axion mass to be negligible at production.

It is straightforward to see that for $\overline f \ll 1$, one recovers the usual formula,
\[
n_\f \sim 2 n_\F \frac{\Gamma_{\F \to \f\f}}{H}.
\]
On the other hand, $\F$ can also decay into other particles via the aforementioned stimulated emission or through very efficient dissipation/nonlinear effects~\cite{Felder:1998vq}, in which case the stimulated decay into axions may be subdominant.\footnote{{One may also consider entropy dilution of the axion abundance from $B=1$, which opens an allowed region for axion production as well. Since in this paper we focus on the radiation-dominated era, we leave such analysis for future work. }}

The energy density of axions produced from $\F$ decay can be estimated generically as
\beq
\rho_\f \sim B\, m_\F^2 M_{\rm pl}^2,
\eeq
where the right-hand side represents the energy density of $\F$ just before decay into axions, i.e. around the onset of oscillation.  
Here, $B (\leq 1)$ again denotes the fraction of the $\F$ energy transferred into axions.\footnote{$B$ may also include the theoretical uncertainty in the initial energy density soon after the onset of $\F$ oscillations. The energy density depends on the initial field value $\F \sim M_{\rm pl}$, the size of $\x_{1}, \x_2$, as well as possible higher-order terms.}

\begin{figure}[!t] 
    \begin{center}
        \includegraphics[width=82mm]{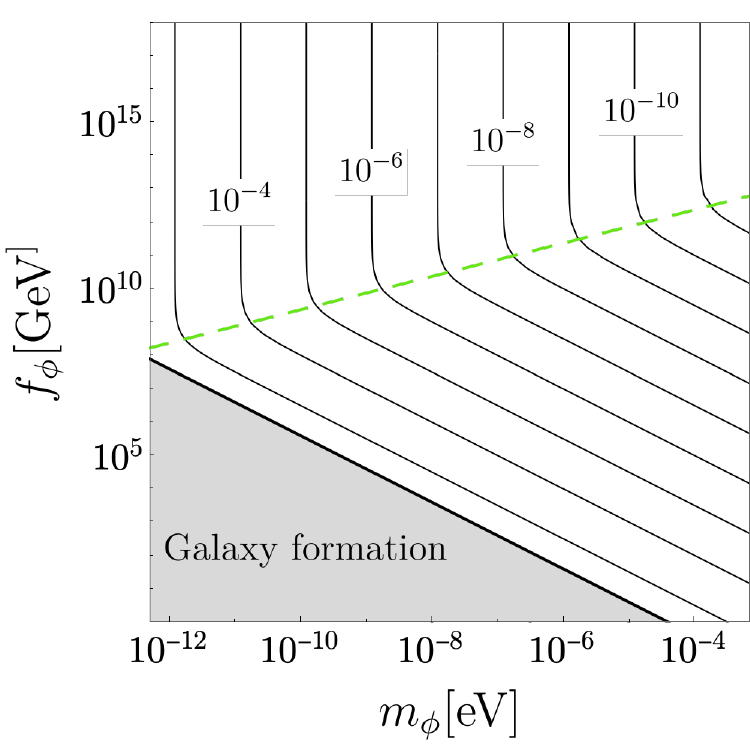} 
        \hspace{1mm}
        \includegraphics[width=82mm]{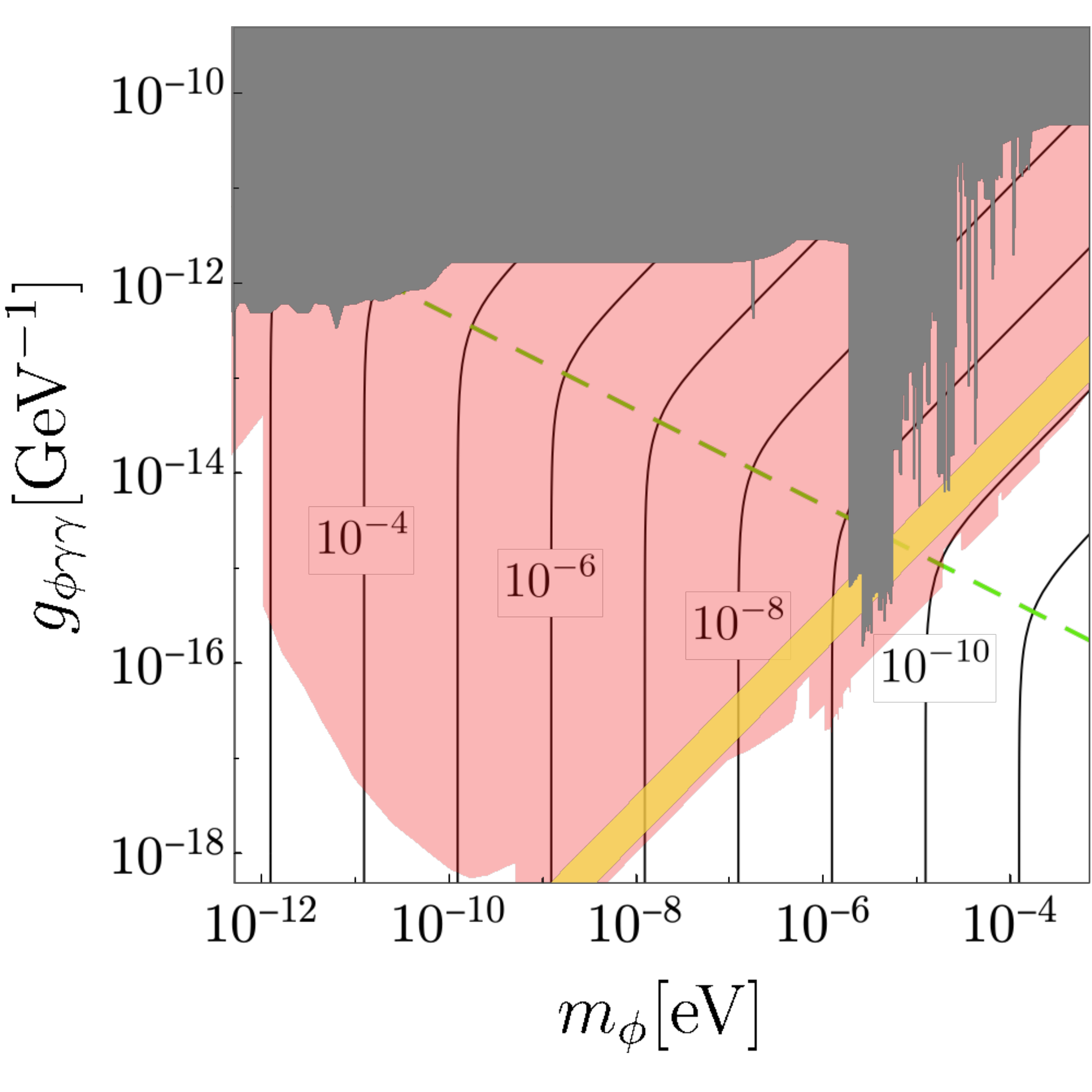}
        \\
\vspace{-5mm}
    \end{center} 
    \caption{In the left panel, we show contours of $B$ in $m_\f \,\text{-}\, f_\f$ plane for light axion dark matter production from particle decay, {while the right panel displays the corresponding contours of $B$ in $m_\f \,\text{-}\, g_{\f\gamma\gamma}$ plane. In both figures,} we take $T_{\rm prod}=10^3\GEV$. {The green dashed line represents the boundary that separates the parameter regions with or without a nonlinear transition, as determined by Eq.~\eqref{eq: Bcondition}.} The lower light-gray shaded region {in the left panel} is excluded due to the generic limit given in \cite{Narita:2025jeg} (see also \cite{Marsh:2019bjr,Dror:2020zru}). {The dark-gray, red, and yellow regions in the right figure correspond to those shown in Fig.~\ref{fig:1DM}.}}
    \label{fig:2DM} 
\end{figure}

Note that the entropy density of the Universe is determined around $m_\F \sim H$, and one can immediately estimate the condition for $B$ using \Eq{ODM} to explain the dark matter:
% \beq
% \{B, B'\} \;\approx\; 
% \Biggl\{2.46 \times 10^{-29}\, g_{s,\star}\, g_\star^{-1/2} \frac{T}{m_\f}, \quad
% 1.23\times 10^{-13}\, g_{s,\star}^{4/3}\, g_\star^{-1} \frac{\GEV^{4/3}}{f_\f^{2/3} m_\f^{2/3}}
% \Biggr\}\Bigg|_{T=T_{\rm prod}},
% \eeq
\begin{align}
    B &\approx 1.2 \times 10^{-9} \, g_{s,\star}\, g_\star^{-1/2} \frac{T}{10^3 \GEV} \frac{10^7 \EV}{m_\phi}, \\
    B' &\approx 4.7 \times 10^{-8} \, g_{s,\star}^{4/3} \, g_\star^{-1} \left( \frac{10^9 \GEV}{f_\phi} \right)^{2/3} \left( \frac{10^7 \EV}{m_\phi} \right)^{2/3}.
\end{align}
where $B^{(')}$ is obtained from $\Omega^{(')}_{\f}=\Omega_{\rm DM}$, respectively. {We also consider the parameter region in which the nonlinear transition occurs. Recalling Eq.~\eqref{eq: pcondition}, the corresponding condition on the parameter $B'$ is given by
\begin{equation}
    \label{eq: Bcondition}
    B' \gtrsim \frac{1}{16} \left(\frac{m_\Phi f_\phi}{m_\phi M_\text{pl}} \right)^2.
\end{equation}
}

Fig.~\ref{fig:2DM} shows contour plots of the effective branching fraction $B$ {for dark matter production} in the $m_\f \,\text{-}\, f_\f$ plane (left panel) and the $m_\phi \,\text{-}\, g_{\phi\gamma\gamma}$ plane (right panel), assuming $T_{\rm prod}=10^3\GEV$. Then, the mass of $\Phi$ is around $10^{-3} \EV$.
To continuously interpolate between the two regimes, we plot $\sqrt{B^2+B'^2}$.  
Note that since $m_\F \sim H$ {at the time of axion production}, the momentum-to-temperature ratio is given by
$
\frac{\overline{p}_\f}{T} \sim \frac{m_\F}{2 T_{\rm prod}} \approx \frac{T_{\rm prod}}{M_{\rm pl}} \sim 10^{-16},
$ 
which is extremely small -- a characteristic feature of this scenario.  
The lower end of the $m_\f$ range is imposed to avoid the region without nonlinear transition, which would otherwise lead to an excessively long free-streaming length,~e.g., \cite{Moroi:2020has,Nakayama:2021avl}. In contrast,  
the upper end is set by the kinematic condition $m_\F > 2 m_\f$.
{In drawing Fig.~\ref{fig:2DM}, we therefore restrict the parameter region of $m_\phi$ to lie within this range. The light-gray region is identical to that in Fig.~\ref{fig:1DMfphi}, but it is no longer effective in the left panel of Fig.~\ref{fig:2DM}, where the astrophysical constraints are more stringent.}

Interestingly, this scenario can also successfully produce a light QCD axion.  
The QCD axion parameter region, which will be discussed in \Sec{QCD axion}, must be analyzed taking into account the temperature effects.  

We emphasize that there exist various possible scenarios other than axion production from modulus-stimulated decay where our discussion is relevant.  
For instance, one may consider axions originating from topological defects or from phase transitions~\cite{Saikawa:2024bta,Kim:2024wku,Buschmann:2024bfj,Benabou:2024msj,Kim:2024dtq,Gorghetto:2018myk,Gorghetto:2020qws,Sikivie:1982qv,Vilenkin:1982ks,Harari:1987ht,Davis:1986xc,Dine:2020pds,Hindmarsh:2021zkt,Saikawa:2024bta,Nakagawa:2022wwm,Lee:2024oaz,Yin:2024txg,Yin:2024pri}.
 Alternatively, we can consider the axion from the very weakly coupled the stimulated decay of Higgs boson~\cite{Nakayama:2021avl,Yin:2024txg}, that is relevant to the symmetry for the axion.

\section{Precise study of axion spectrum: minicluster formation and gravitational waves}
\lac{3}

The nonlinear transition is accompanied by short-lived domain walls.\footnote{See also recent scenarios for domain wall formation: from the axion roulette, i.e., an axion rotation around the periodic direction via multi-axion mixings and level crossing~\cite{Daido:2015bva, Daido:2015cba}, from inflationary fluctuations with robust stability against the population bias~\cite{Gonzalez:2022mcx}, and from negative thermal mass effects~\cite{Yin:2024txg,Yin:2024pri}. None of these require the symmetric phase usually associated with domain walls.}
These domain walls form because the field fluctuations of the axion before the transition are so large that the axion fields at different places settle to different vacua in the periodic potential.  

Although a scaling solution of domain walls can emerge, population bias causes the collapse of the domain wall network within a few Hubble times~\cite{Vilenkin:1981zs,Sikivie:1982qv,Mohanty:1984pj,Gelmini:1988sf,Lalak:1993ay,Lalak:1993bp,Lalak:1994qt,Coulson:1995nv,Coulson:1995uq,Larsson:1996sp,Correia:2014kqa,Correia:2018tty,Krajewski:2021jje}. 
(See also Refs.~\cite{Gonzalez:2022mcx,Kitajima:2023kzu} for studies of the stability of such networks against population bias in the case of inflationary fluctuations, which we do not assume here in order to avoid the isocurvature bound.)
Those features are highly nonlinear in fields, and we solve the system with lattice simulation.

To study the system precisely we modify the {\tt Cosmolattice}~\cite{Figueroa:2020rrl,Figueroa:2021yhd} with the following feature. 
The initial fluctuation $\d \f^0 $ satisfies (for a similar setup see Ref.\,\cite{Narita:2025jeg})
\beq
\langle \delta \f^0 \delta \f^0 \rangle =  \int \frac{d k}{k} \, {\cal P}_{\delta \f}(k).
\eeq
The reduced power spectrum is taken 
\beq
{\cal P}_{\delta \f}(k) = \Theta[K_{\rm UV}-k] \tilde{\phi}^2 \(\frac{k}{m_\f}\)^3,
\eeq
with $\tl \f$ being a normalization, $\Theta$ the Heaviside step function, and 
$K_{\rm UV}$ the momentum cutoff for the spectrum we consider.
This is the form of white noise in the large scale and the isocurvature mode is highly suppressed. $K_{\rm UV}$ is supposed to be smaller than $m_\f$ so that we only have non-relativistic modes. 
%One can immediately see that when the $\vev{\d\f^0\d\f^0}\gg f_\f^2 $

%Then we get
%\beq
%\langle \delta \f^0 \delta \f^0 \rangle= \frac{ \tilde{\phi}^2 }{3}  (\frac{K_{\rm UV}}{m_\f})^3,
%\eeq
%while the $q\to0$ limit gives the case for $q=0$.

\begin{figure}[!t] 
    \begin{center}
        \includegraphics[width=155mm]{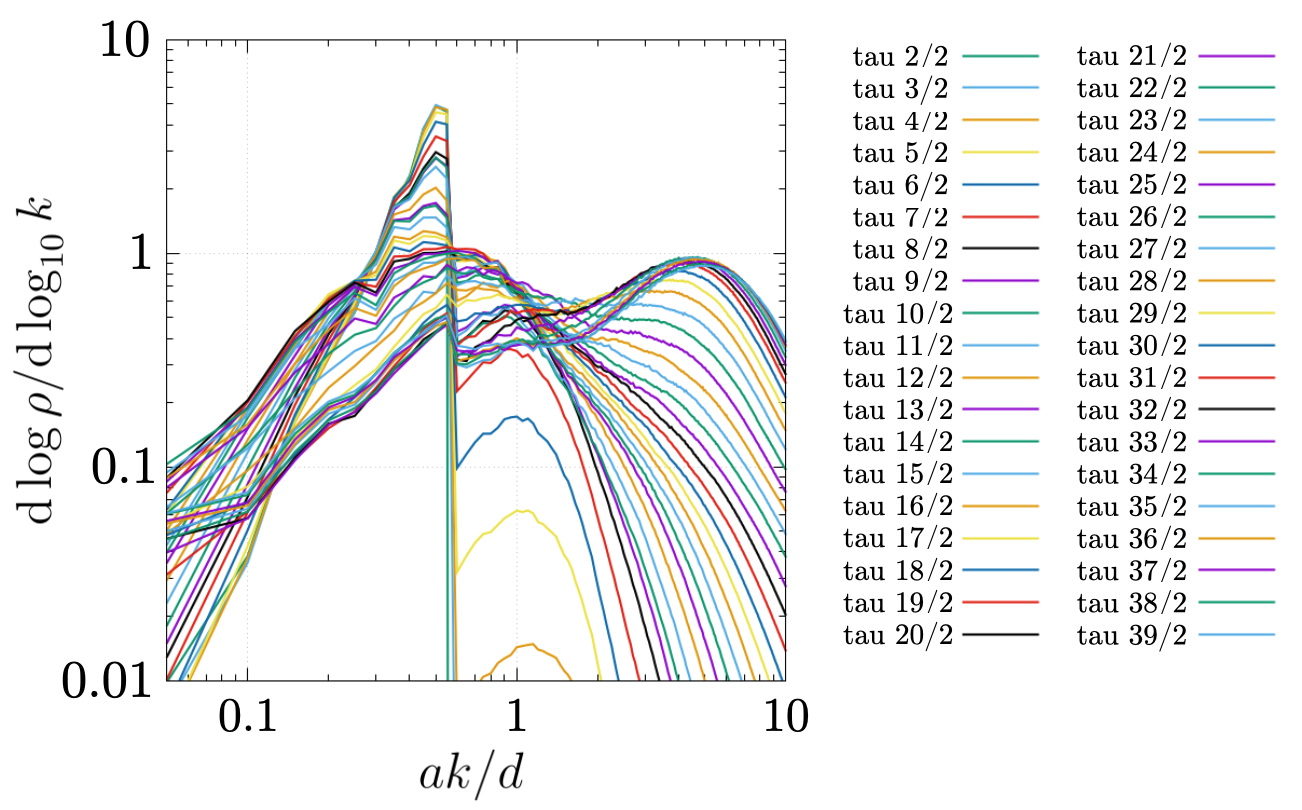}
        \vspace{0mm}
    \end{center} 
    \caption{Lattice simulation result for $\delta[k]$.  
    The distribution flattens in time with a peak around the Hubble scale. $\rm tau$ denotes $\tau d$ with $\tau$ being the conformal time. We take $\{m_\phi^2,K_{\rm UV}^2 ,\tilde{\phi}\}=
\{0.8d^2,0.4m_\phi^2,8d\}$}
    \label{fig:3} 
\end{figure}

\subsection{Evolution of the system from axion spectra, and axion clumps}

To study the axion spectrum, we use the lattice simulation to check the overdensities of the axion soon after the nonlinear transition in Fig.\ref{fig:3} for $1024^3$
%$512^3, 1024^3, \AND 512^3,$ 
 lattice simulation with $\{m_\phi^2,K_{\rm UV}^2 ,\tilde{\phi}\}=
% \{0.3d^2,0.9m_\phi^2,8d\},
\{0.8d^2,0.4m_\phi^2,8d\}
%,
%\AND\{0.8d^2,0.9m_\phi^2,8d\}
. $
%  respectively. 
  We take the $2\pi/L=0.05 d$ with $L$ being the box size, $f_\f=d$, and $H_0=0.5 d$ with $d$ being the machine unit in all the figures.\footnote{We cannot separate the scales significantly for the stability of the simulation.}

In these figures we evaluate the so-called overdensity parameter (or the normalized axion spectrum) 
\beq
\delta[k]\equiv {\partial _{\log k} \log\rho_\f}.\\
\eeq
in different conformal time $\tau=(1,1.5,\cdots 19., 19.5)d^{-1}$. 
Here $\partial _{\log k}\rho_\f$ is defined by the $\frac{k^3}{(2\pi)^3}\int d\Omega {\f_{\vec{k}} \f_{-\vec{k}} (k^2+\vev{\partial_\f^2 V})}$ with $k=|\vec{k}|$ fixed. 
Strictly speaking we plot $\log{10}\times \delta[k]$. 

We carefully choose the parameter so that the short lived domain walls collapse enough. 
%Although the left panel has lower initial momentum than the right panel, the final distribution shows a peak around the same position which is $\lesssim 2\pi H_0/\O(1).$ Note that transition happens at $m_\f^2 f_\f^2\sim k_{UV}^2 \tl{\f}^2$ which does not change. 
At the beginning around $\tau \sim 10/d$, the spectrum is not relevant to the initial modes but it is relevant to the transition scale Hubble parameter, which is confirmed by changing the parameters. This can be understood from the formation of the domain walls~\cite{Narita:2025jeg}, because the domain walls have the typical curvature around the Hubble parameter $H$. Indeed, the shape of $\delta$ is similar to the power spectrum of $\phi$ with scaling domain wall, which were studied in the context of cosmic birefringence~\cite{Takahashi:2020tqv,Kitajima:2022jzz, Gonzalez:2022mcx,Kitajima:2023kzu}. 

However, afterwards, around $\tau\sim 20 d^{-1}$ the dominant mode becomes much higher. This can be understood from the collapse of the domain wall network  due to the population bias. Then the semi-relativistic axions are produced. 
The momentum is around the mass, and we confirm \Eq{ptr}.
%\beq
%\overline{p}_\f \sim m_\f.
%\eeq
%Therefore 

The collapses of topological defects are known to induce axion clumps such as axion miniclusters and axion stars~\cite{Hogan:1988mp,Kolb:1993zz,Kolb:1993hw,Kolb:1994fi,Kolb:1995bu} (see also more recent simulations in Refs.~\cite{Vaquero:2018tib,Buschmann:2019icd,Ellis:2020gtq,Eggemeier:2019khm,Ellis:2022grh}). In our scenario we expect the same conclusions. 
Given $\delta(k)$, one can estimate the formation and its distribution following these references.

%In the middle panel, we show the case with smallest momentum. The condition of $m_\f^2 f_\f^2\sim k_{UV}^2 \tl{\f}^2$ is satisfied earlier than the other two. One can see although the typical initial momentum is the same as the left panel, the final peak comoving momentum is slightly shifted in the high momentum mode, which corresponds to the Hubble at earlier stage. 
% with $k \sim H_{\rm tr},$ which is the typical scale of the domain wall separation, which is irrelevant to the cutoff of the initial fluctuations.

%where $\overline{p}_\f^{\rm pre}$ is the typical momentum soon before the nonlinear transition. 

%Since axion clumps are likely to form, their effects may be relevant for axion search experiments.  

\subsection{Gravitational waves}

\begin{figure}[!t] 
    \begin{center}
\includegraphics[width=155mm]{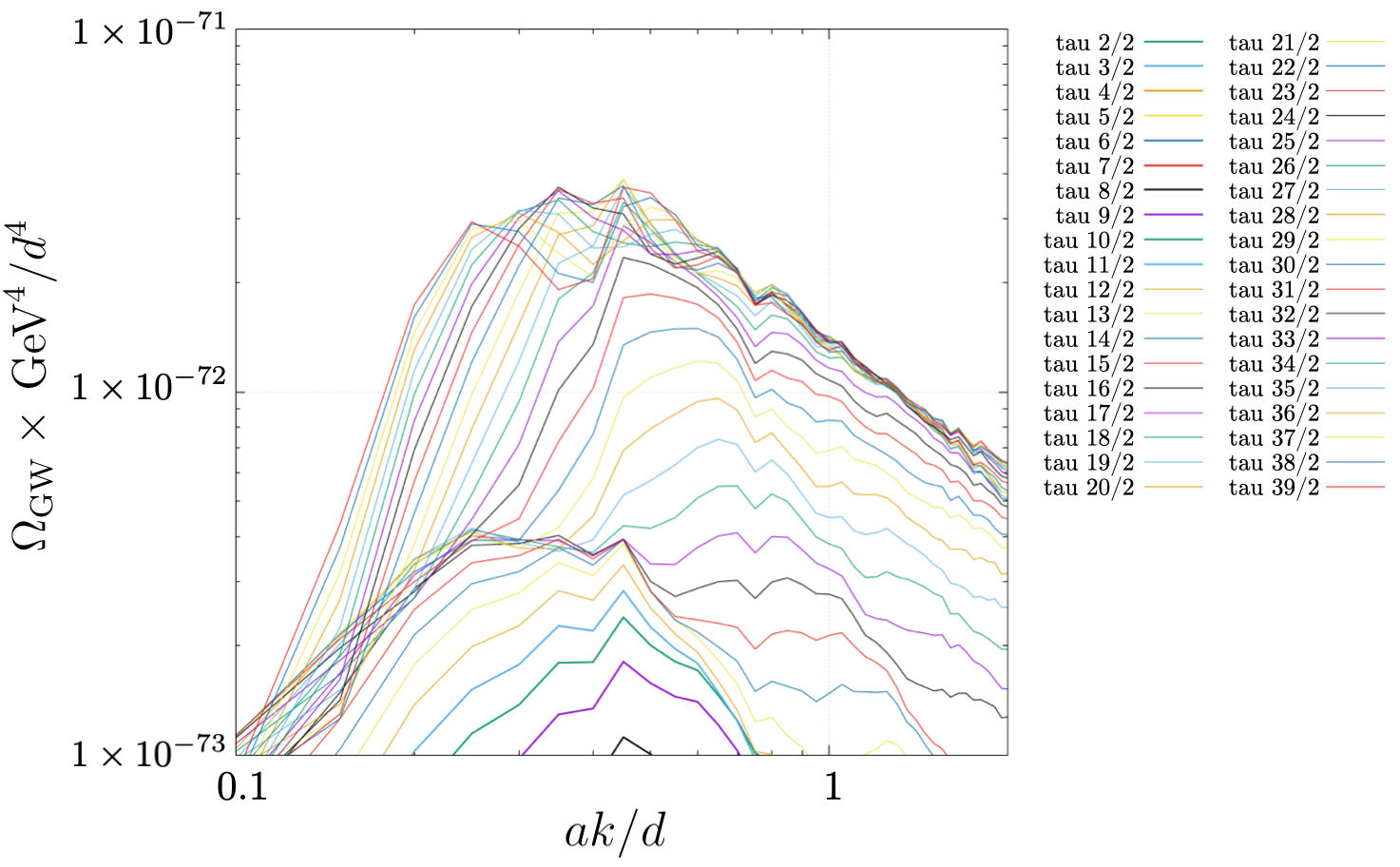}
        \vspace{0mm}
    \end{center} 
    \caption{Gravitational wave emission in terms of $\Omega_{\rm GW}\times \GEV^4/d^4$ for the right panel of Fig.~\ref{fig:3}.}
    \label{fig:4} 
\end{figure}

The nonlinear transition can induce gravitational waves.  
In particular, the collapse of short-lived domain walls is expected to produce a stochastic background of gravitational waves.  
For this contribution, the peak frequency corresponds to the typical curvature scale of the short-lived domain walls, which is of order $H_{\rm tr}$.  
Thus, the comoving peak momentum is
\beq
a k_{\rm peak} \sim a H_{\rm tr}.
\eeq
Similarly, the gravitational wave spectrum for domain wall collapse can be estimated following \cite{Hiramatsu:2013qaa} as
\beq
\Omega^{\rm DW}_{\rm gw,peak} \sim 
% \left.
\frac{\sigma^2}{24\pi M_{\rm pl}^4 H_{\rm tr}^2}
% \right|_{T=T_{\rm tr}}.
\eeq
Here $\sigma=8 f_\phi^2 m_\f$ is the domain wall tension.

This estimate can be checked numerically, as shown in Fig.~\ref{fig:4}, which corresponds to Fig.~\ref{fig:3}. In Fig.~\ref{fig:4}, we use $512^3$ lattices for reducing the calculation cost. We have checked that the $\delta(k)$ looks similar except for the modes with $a k/d>10$.  
By taking $\sigma = \sqrt{0.8}\, d^3$ and $H \sim 0.05\, d$, i.e. $\tau \sim 10/d$, which is soon before the domain wall collapse for the parameter in the figure, one obtains 
$
\Omega^{\rm DW}_{\rm gw,peak}\sim 10^{-71}\, d^4/\GEV^4
$
from the analytic formula, which is consistent with the numerical result, which includes the enhancement from the collapsing effect found in \cite{Kitajima:2023cek}. 

To match the gravitational wave spectrum today, we must take into account the redshift, which introduces an additional suppression of $\sim 10^{-5}$. 
Thus, the resulting gravitational wave abundance is too small to be observed, since
$
10^{-5}\,\Omega_{\rm gw,peak}^{\rm DW} \lesssim 2\times 10^{-12}\left(\frac{\KEV}{T_{\rm tr}}\right)^2 .
$
Here the r.h.s.\ is obtained by solving $\Omega' = \Omega_{\rm DM}$ with $m_\phi = H_{\rm tr}$, so that the domain wall tension at a given $T_{\rm tr}$ is maximized. 
This suppression arises because the domain wall density is reduced by the requirement that the axion account for dark matter, which remains subdominant during the radiation-dominated epoch. 

\section{QCD axion}
\lac{QCD axion}

So far, we have focused on the case where the decay constant and the mass do not vary with time or temperature.
For the decay constant, this assumption is natural if $f_\f^4 \gg \rho_\f$ in the relevant epoch.\footnote{Here, we assume that the Higgs field responsible for axion symmetry breaking has a scale as large as $f_\f$.  
In the case of a weakly coupled theory, this argument does not hold, and $f_\f$ may evolve with time even at late epochs~\cite{Yin:2024txg,Yin:2024pri}.  
Such scenarios are interesting possibilities for future study.}  
On the other hand, a time-dependent mass of the axion is natural, particularly for the QCD axion that solves the strong CP problem.  

The topological susceptibility of QCD, $\chi(T)$, depends on temperature as
\begin{align}
    \chi(T) \simeq
    \begin{cases}
        \chi_0 & (T < T_\mathrm{QCD}), \\[6pt]
        \chi_0 \left( \dfrac{T}{T_\mathrm{QCD}} \right)^{-n} & (T \geq T_\mathrm{QCD}),
    \end{cases}
\label{eq:chi}
\end{align}
where we adopt $\chi_0 = (75.6\,\mathrm{MeV})^4$, $T_\mathrm{QCD} = 153\,\mathrm{MeV}$, and $n = 8.16$~\cite{Borsanyi:2016ksw}.  
The axion mass is then determined by
\beq
m_\f^2(T)\, f_\f^2 = \chi(T).
\eeq

Since the topological susceptibility varies slowly {with time, the adiabatic approximation allows us to apply} the previous dynamics with the temporary potential. Given that $\chi(T), m_\f(T)$ increases in time {as the Universe cools down}, the condition for the nonlinear transition to happen is
\beq
\rho_\f > \chi(T) \text{~when~} \overline p_\f \sim m_\f(T).
\eeq
After the transition $\rho_\f \sim \chi(T_{\rm tr}),$ the comoving number begins to be conserved. 
Then, we can estimate the abundance as 
\beq
\Omega_\phi^{'} \sim  m_\f(T\to 0) \left.\frac{\chi(T)}{s m_\f(T)}\right|_{T=T_{\rm tr}}\frac{s_0}{\rho_c}.
\eeq
The matching with the dark matter abundance $\Omega_\phi^{'}=\Omega_{\rm DM}$ gives 
\beq
T_{\rm tr}\sim 1.26\GEV.
\eeq
This is close to the temperature for the onset of oscillation in the misalignment mechanism because the condition is close $\chi(T)\to \chi(T) \theta_i^2$ with $\theta_{i}$ being initial misalignment, which is naturally unity. With $\theta_{i}=1$, the correct dark matter abundance leads to $f_\f\sim 10^{12}\GEV, m_\f\sim 5.7\times 10^{-6}\EV,$ which corresponds to $m_\f(T_{\rm tr})\sim H_\text{{tr}}.$ 

In our case, we need $m_\f(T_{\rm tr})(>\overline{p}_\f)>H_\text{{tr}}$. Thus, we have 
\beq
f_\f \lesssim 10^{12}\GEV,~~\AND~~ m_\f\gtrsim 5.7\times 10^{-6}\EV.
\eeq
This gives motivation to a relatively heavy axion: $10^{-6}\EV<m_\phi<0.1 \EV$ where the upper limit is from the star cooling bounds ($f_\f\gtrsim 10^8\GEV$).

However, in this case, we need the axion to have low momentum before the nonlinear transition. For instance,  with $f_\f\sim 10^8\GEV$, $\overline{p}_\f/T_{\rm tr}<10^{-14}$, this is achieved for the sample parameter taken in Fig.\ref{fig:2DM} for the moduli-stimulated decay scenario.

\section{Conclusions and discussion}
\lac{6}

In this work we revisited axion/ALP dark matter scenarios in which the axion is produced with non-zero momentum rather than through coherent misalignment.  
The central input is the dynamical criterion for the axion to behave as non-relativistic matter, \Eq{1}, which refines the usual requirement $\overline p_\phi < m_\phi$ by demanding that the potential energy scale $m_\phi^2 f_\phi^2$ is already dominant.  
When this condition is violated at the epoch $\overline p_\phi \sim m_\phi$, the axion fluid effectively behaves as radiation and the comoving {\it number} is not conserved across the ensuing nonlinear transition; instead, the comoving {\it energy} is approximately conserved.  
We formulated a practical prescription to compute the relic abundance in this regime and showed that the standard estimate based on comoving number conservation must be replaced by the ``transition-anchored'' estimate controlled by $T_{\rm tr}$ defined via $\rho_\phi(T_{\rm tr}) = m_\phi^2 f_\phi^2$.  
In particular, in the simple string-inspired scenario where axions are produced through stimulated modulus decay, this parameter region becomes especially important when $f_\phi$ is relatively small, the regime favored by experimental axion searches.  
In other words, when targeting axion dark matter in experimentally accessible regions, one must carefully account for the nonlinear transition in cosmological studies.  

A second outcome of our analysis is phenomenological: when $\rho_\phi$ redshifts down to $m_\phi^2 f_\phi^2$, large field inhomogeneities generically form short-lived domain walls which collapse within a few Hubble times.  
This episode seeds $\mathcal{O}(1)$ overdensities on the mass scale at $T_{\rm tr}$ and therefore provides natural initial conditions for axion clumps such as miniclusters and axion stars.  
The same dynamics also source a stochastic gravitational-wave background whose peak frequency and amplitude are set by $H_{\rm tr}$ and the transient wall tension, although the signal is too small to be observed in the foreseeable future.  

This suppression does not occur if the axion is heavy and does not constitute dark matter.  
In the case of a very heavy axion,  the nonlinear transition can instead induce significant gravitational waves.  
In this scenario, one needs to suppress the remnant axion entropy production, which can be easily achieved (see appendix \ref{chap:52}). The primordial blackholes can be formed which can instead form the dark matter.

\section*{Acknowledgement}
This work is supported by JSPS KAKENHI Grant Nos. 22K14029 (W.Y.), 22H01215 (W.Y.), Graduate Program on Physics for the Universe (Y.N.), and JST SPRING, Grant Number JPMJSP2114 (Y.N.). W.Y. is also supported by Selective Research Fund from Tokyo Metropolitan University.

 \appendix 
 \section{Nonlinear transition parameter region with large decay constants}

\lac{5}

So far, we have shown that the nonlinear transition can be important in axion dark matter scenarios. The predicted axion decay constant is not very large. Thus, the axion may originate from quantum field theory or from a string axiverse with large-volume compactifications.  
We now study the conventional string axion with large decay constants $10^{15-17}\GEV$.

Suppose again $\Phi \to \phi\phi$, where $\phi$ is the axion, with the Lagrangian \eq{Lag}.
Around the onset of oscillation of $\Phi$, we have $\rho_\Phi \sim m_\Phi^2 M_{\rm pl}^2/2$. On a similar timescale, $\phi$ is produced from stimulated emission with, for simplicity, $B \approx 1$, i.e.
\beq 
\rho_\phi \sim \rho_\Phi \sim m_\Phi^2 M_{\rm pl}^2/2.
\eeq
 Note that since $m_\Phi \sim H$ is the condition for the onset of oscillation, the energy density is close to the total energy density of the Universe. Here we assume again for simplicity that the modulus completely disappears due to some other interactions. 
 % Removing this assumption does not change the conclusions in this part.  
% In general, some remnant of $\Phi$ may remain depending on the coupling and mass hierarchy (see Eq.\,\eqref{number}), which may later dominate the Universe and reheat it. 

At the time of stimulated production, the axion has an energy density, $m_\F^2 M_{\rm pl}^2/2$, larger than $m_\phi^2 f_\phi^2$. The typical axion momentum $\bar{p}_\phi \sim m_\Phi/2$ redshifts to $m_\phi$ with an expansion $a(T_{m_\phi})/a(T_{\rm prod}) \sim m_\Phi/(2m_\phi)$. We then obtain the axion radiation energy density $(2m_\phi/m_\Phi)^4 \times m_\Phi^2 M_{\rm pl}^2/2$. Therefore, from Eq.\,\eqref{eq: cond1} we find that a nonlinear transition occurs if
\beq \laq{heavyaxionrange}
m_\phi \gtrsim m_\Phi \frac{f_\phi}{\sqrt{8} M_{\rm pl}}. 
\eeq

Taking $f_\phi = 10^{15-17}\GEV$, we then see that the axion relevant for the nonlinear transition should satisfy
\beq 
(10^{-4}-10^{-2})m_\Phi \lesssim m_\phi < 0.5m_\Phi .
\eeq  
This implies that for nonlinear transitions with large decay constant, the redshift from the modulus oscillation to the nonlinear transition is not very large.\footnote{This implies even if the moduli remains, it is natural to assume that the remnant of $\Phi$ remains subdominant at the transition or that the energy density is not much higher than the radiation. The  coherent oscillation of the modulus remnant also behaves as nonrelativistic matter later and reheat the Universe. The reheating temperature is similar to that for the string ALP. }

The nonlinear transition occurs with an energy density comparable to the total energy of the Universe.\footnote{This is a generic prediction if after the moduli dominant stimulated decay into the axion, one has radiation dominated Universe.} 
After the nonlinear transition, nonrelativistic axion particle production arises from the collapse of the domain wall network. 
The component dominates the Universe and later reheats it through couplings to Standard Model particles. 
Assuming the axion is an ALP, the reheating temperature can be larger than MeV, sufficient for successful big bang nucleosynthesis, if 
\beq 
m_\Phi> 0.5 m_\phi \gtrsim 1-10^3\TEV ,
\eeq 
as in the usual solution to the moduli problem.

\section{Primordial black hole formation from nonlinear transition of heavy axion}
\lac{52}
The short-lived domain wall formation generates overdensities on the Hubble scale, as shown in Fig.~\ref{fig:3}. If $B\approx 1$, and the modulus completely disappears after the stimulated emission into axions, the axion is a dominant component of the Universe around the transition, and the overdense region in Fig.~\ref{fig:3} can be approximated as the ratio 
\beq 
\delta(k) \sim \partial_{\log k} \rho_\phi / \rho_{\rm tot},
\eeq
with $\rho_{\rm tot}$ being the total energy density of the Universe.

This overdensity in real space, obtained from the Fourier transformation, can be as large as the horizon size (at least for $\tau \sim 10/d$ in Fig.~\ref{fig:3}), and can dominate the Universe. 
% In other words, the domain wall loops typically have the size of the Hubble scale, which leads to overdensities. 

The overdense regions collapse into PBHs with mass
\beq 
M \sim \frac{4\pi}{3}\left(\frac{2\pi}{k}\right)^{3} 
\left.\partial_{\log k}  \rho_{\phi}/\rho_{\rm tot}\right|_{k=H_{\rm tr}},
\eeq 
within $\mathcal{O}(1)$ Hubble time, 
before the population bias becomes strong enough to annihilate the domain walls. 
Thus we expect PBH production. For instance, for the large decay constants, we have 
\beq 
M\sim 10^{-55}M_{\odot} \(\frac{f_\f}{10^{15}\GEV} \frac{m_\f}{10^{4}\GEV}\)^{-3}
\eeq 
which will soon evaporate.
On the other hand, we have 
 PBHs of mass $10^{-16}M_{\odot}$ 
 for $f_\f=10^4\GEV, m_\f=100\GEV$, which could be the dominant dark matter, although the axion, if it couples to the Standard Model particles, has a very short lifetime.  In this case, we can even have the strong gravitational wave with the frequancy in the target range of DECIGO, LISA etc. 

Note that this mechanism differs from PBH formation via potential bias, in which the false vacuum energy is also important, e.g.,~\cite{Ferrer:2018uiu,Kitajima:2023cek,Gouttenoire:2023ftk}. 
Even in the potential-bias case, estimating the PBH abundance precisely is a difficult task. 
Therefore, we do not attempt to evaluate the PBH abundance in detail here. We note that the PBH abundance can be easily reduced by considering $B<1$ and since this lowers the fraction of domain wall energy in the total energy density, making overdense regions rarer.

 \bibliography{GenericALPDMbound.bib}
\end{document}